\documentclass[english,notitlepage]{revtex4-2}
\usepackage[T1]{fontenc}
\usepackage[latin9]{inputenc}
\usepackage{wrapfig}
\usepackage{amsbsy}
\usepackage{amstext}
\usepackage{amssymb}
\usepackage{graphicx}
\usepackage{esint}

\makeatletter
\usepackage{hyperref}

\makeatother

\usepackage{babel}
\begin{document}
\title{Instabilities, Anomalous Heating and Stochastic Acceleration of Electrons
in Colliding Plasmas}
\author{M.A. Malkov$^{1}$ and V.I. Sotnikov$^{2}$}
\affiliation{$^{1}$CASS and Department of Physics, University of California, San
Diego, La Jolla, CA 92093\\$^{2}$Air Force Research Laboratory,
WPAFB, OH}
\begin{abstract}
The collision of two expanding plasma clouds is investigated, emphasizing
instabilities and electron energization in the plasma mixing layer.
This work is directly relevant to laboratory experiments with explosively-created
laser or z-pinch plasmas but may also elucidate naturally occurring
plasma collisions in astrophysical or space physics contexts. In the
previous publications \citep{MalkovSotnikov2018PhPl,Sotnikov2020},
we have studied, analytically and numerically, the flow emerging from
interpenetrating coronas launched by two parallel wires vaporized
in a vacuum chamber. The main foci of the studies have been on the
general flow pattern and lower-hybrid and thin-shell instabilities
that under certain conditions develop in the collision layer. The
present paper centers around the initial phase of the interpenetration
of the two plasmas. A two-stream ion-ion instability, efficient electron
heating, and stochastic acceleration dominate plasma mixing at this
phase. Both the adiabatic (reversible) electron heating and stochastic
(irreversible) acceleration and heating mechanisms, powered by unstably
driven electric fields, are considered. The irreversibility results
from a combination of electron runaway acceleration in the wave electric
fields and pitch-angle scattering on ions and neutrals.
\end{abstract}
\maketitle

\section{Introduction}

Macroscopic plasma motions often harbor microscopic phenomena that
are challenging to understand or even observe. The hi-speed collision
of two plasma clouds is an excellent example of how microinstabilities
control the macroscopic flow that results from the collision. The
collision of ordinary gases results in a contact surface and a pair
of shocks propagating into each gas. No mixing other than a relatively
slow diffusion across the contact surface is expected, at least in
a perfectly symmetric stable flow. Plasma collisions are different
in several ways. First, colliding plasma clouds may go through each
other without much effect. This regime is expected when the relative
plasma motion is collisionless, owing to its high velocity, whereas
the individual plasmas may or may not be. At the same time, instabilities
associated with the relative motion (two-stream instability) may have
no time to develop during the plasma interaction or be suppressed
by the internal pressure of the streams \citep{VVS_1961SvPhUsp}.
The difference in collisionality regimes may occur when the relative
velocity between the clouds is much larger than the thermal velocity
in each of them, owing to the steep growth of the collision mean free
path with the relative particle velocity, $\lambda\propto v^{4}$.
Sustainable counterstreaming ion flows have also been observed in
numerical simulations with a contact discontinuity-type of initial
state \citep{Moreno_2019}. The interpenetration of adjacent plasmas
of different densities results in a phase space structure where mixed
populations intersperse residual areas of counterstreaming protons.
So the phase-space develops density holes.

This paper will consider microscopic instabilities and electron heating
in interpenetrating plasmas using unmagnetized plasma approximation
near a symmetry plane of collision. The collision regime of magnetized
electrons and unmagnetized ions has been addressed in \citep{MalkovSotnikov2018PhPl}
along with the macroscopic two-dimensional flow emerging from the
collision. Macroscopic instabilities of this flow have been studied
numerically in \citep{Sotnikov2020} in hydrodynamic approximation.
When focusing on the microscopic instabilities and associated heating
processes, further progress can be made by considering plasma interpenetration
along the centerline in one dimension, which we pursue in this paper.

Plasma collisions are best studied in experiments with laser and z-pinch
plasmas. The present work is motivated by laboratory experiments with
two ohmically-exploding parallel wires. Being vaporized, they launch
plasma coronas toward each other. The interpenetration and mixing
of the hot coronas are followed by a collision and mixing of denser
phases of the melted wire material. They are also much colder than
the coronas, weakly ionized, and move at an order of magnitude lower
speed (typical at 3-5 km/s vs. 50-100 km/s, \citep{Sarkisov2005}).

Nevertheless, this phase is characterized by protracted light emission
from the collision layer, which is considerably less pronounced and
shorter in time in the case of a single wire explosion. The difference
speaks to the importance of the corona interpenetration stage of the
collision process. Otherwise, two closely spaced parallel wires would
exhibit the same flow morphology as a single wire with a similar total
energy deposition. On the contrary, the single- and double-wire explosions
show different emission patterns in intensity and morphology. Namely,
the double-wire explosion develops a thin glowing layer between the
two wires. It extends symmetrically to a chamber scale, remains narrower
than the gap between the wires, and persists for times until their
core material engulfs the entire chamber.

Turbulent electron heating and acceleration during the early stage
of corona interpenetration may enhance the line emission from the
colliding neutral core materials that trail coronas. This premise
and the above considerations have motivated the present study. Besides,
we are interested in characterizing the entire collision in terms
of its scales, growth rates, and thresholds of underlying instabilities.
Spatio-temporal characteristics of the instabilities and associated
turbulent motion is another objective of this study.

The analytic and numerical studies in \citep{MalkovSotnikov2018PhPl,Sotnikov2020}
have indicated that plasmas that stream against each other from two
ohmically vaporized wires may indeed mix in a relatively thin layer,
in accord with the preceding experiments \citep{Sarkisov2005}. The
shocked corona layer is separated from the inflowing material by a
pair of termination shocks (for supersonic flows), and the plasma
pressure inside the layer reaches its maximum at the flow stagnation
point. The emerging pressure gradient drives two symmetric outflows
perpendicular to the plane of the wires. These outflows are squeezed
between the inflowing coronas of the exploded wires (see, e.g., Figs.4-9
in \citep{Sotnikov2020}). A similar flow pattern forms
a dense material of the wires that follows the corona flow (Fig.10
in the above reference). The flow pattern is qualitatively similar
for the collisionless and hydrodynamic regimes. The role of pressure
in the former case plays electrostatic and pondermotive potentials.
The electrostatically mediated flow in the case of Boltzmannian electrons
is formally equivalent to the hydrodynamic flow with an adiabatic
index $\gamma=1$. This analogy justifies hydrodynamic simulations
of colliding plasmas \citep{Sotnikov2020} as a guide for studying
the microscopic processes to which we turn in the present paper. Besides,
the $\gamma=1$ value indicates that the shocks are radiative, which
is consistent with the enhanced emission from a surprisingly thin
shocked plasma layer observed in the experiments. The thin layer points
to high shock compression, characteristic of radiative shocks.

The remainder of the paper is organized as follows. In the next section,
we consider instabilities that we expect to develop when two coronas
from exploded wires begin to penetrate each other. In Sec.\ref{sec:Two-Stream-Instability-in},
we first evaluate the electron response to the unstable waves and
obtain their velocity distribution established in the turbulent electric
field of these waves in an adiabatic (non-resonant) wave-particle
interaction regime. Next, in Sec.\ref{sec:Stochastic-Heating-and},
we include the electron-ion and electron-neutral collisions, thus
making electron dynamics irreversible. In particular, we will obtain
the steady-state analytic solutions to the problem and time-dependent,
numerical ones. We conclude the paper in Sec.\ref{sec:Conclusions-and-Discussion}
with a summary of the results and a brief discussion.

\section{Two-Stream Instability in Counterstreaming Plasmas\label{sec:Two-Stream-Instability-in}}

When two coronas expanding from electrically exploded wires collide
supersonically, the collision region is bound by a pair of termination
shocks that propagate into the inflowing plasmas on each side of the
shocked layer. In both high- and low-Mach-number cases, an electrostatic
potential builds up between interpenetrating coronas and slows their
ions down before the mixed flow deflects off the plane of the two
wires. To produce such deflection, the electrostatic potential, $\Phi$,
must reach the level of

\[
e\Phi\sim Mu^{2}
\]
\begin{wrapfigure}{o}{0.5\columnwidth}%
\includegraphics[scale=0.47]{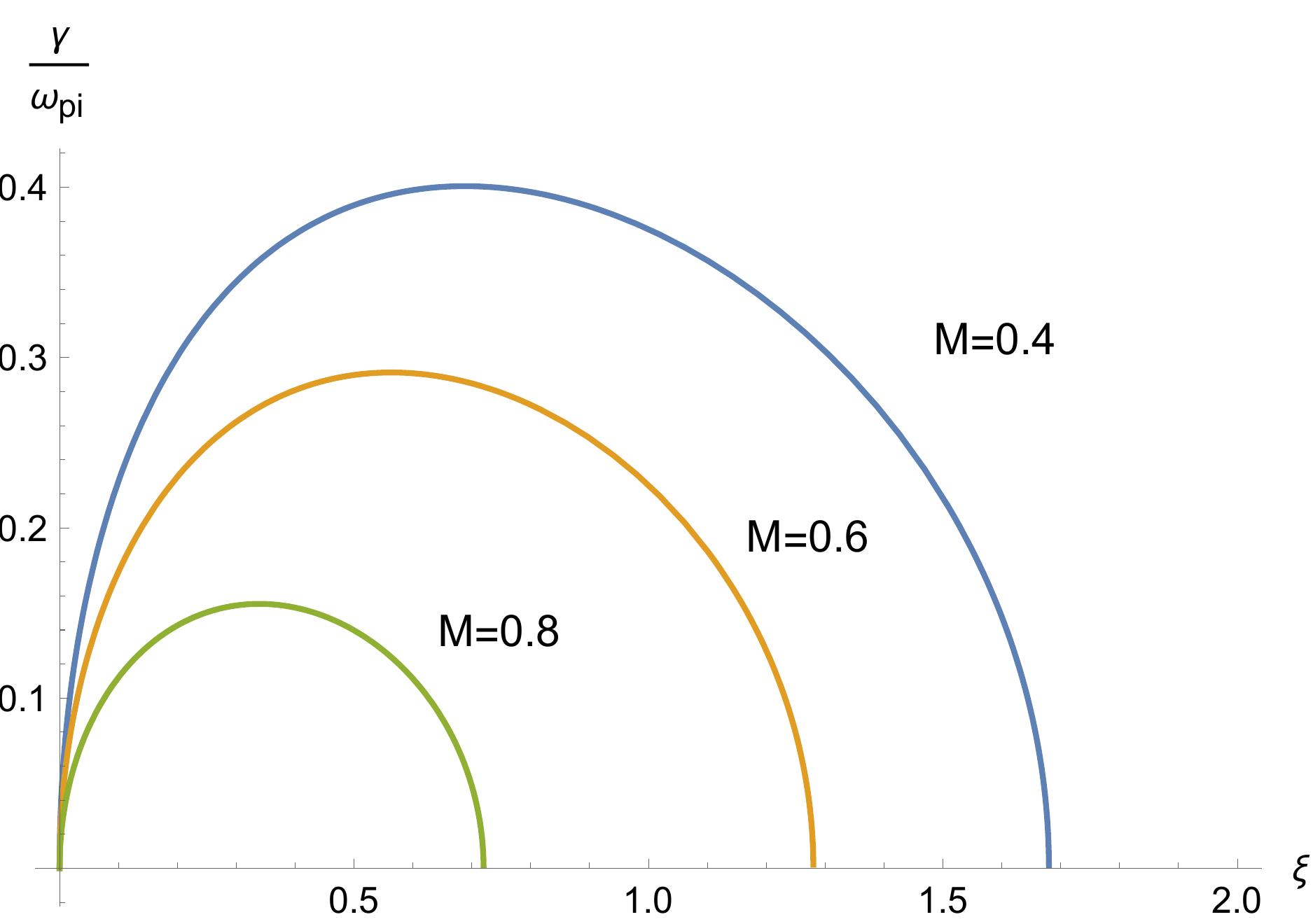}

\caption{Growth rate of the two-stream instability as described by eq.(\ref{eq:GrowthRate2stre}).
\label{fig:Growth-rate-of}}
\end{wrapfigure}%
where $u$ is the speed of plasma upstream of the collision layer,
and $M$ is the ion mass. In more detail, a two-dimensional flow that
emerges from the collision of two concentric outflows has been studied
using the hodograph transformation in \citep{MalkovSotnikov2018PhPl}.
The magnetic field carried to the collision layer with the flows is
assumed to have no substantial effect on the ion motion since the
flow ram pressure is typically about an order of magnitude higher
than the magnetic pressure \citep{Sarkisov2005}. This condition is
equivalent to the ion Larmor radius $\rho_{i}\sim u/\omega_{ci}$
being much larger than the ion skin depth, which is the basic scale
of a magnetized collisionless shock. However, electrons are likely
to be magnetized, at least before being heated, so that the electric
field produces an $\boldsymbol{E}\times\boldsymbol{B}$ drift of electrons
with the drift velocity:

\[
V_{E}\sim c\Phi/LB
\]
where $L$ is the characteristic scale of the potential $\Phi$ that
slows the ions down. Therefore, the electron drift velocity $V_{E}$
can be estimated as follows:

\[
V_{E}\sim\frac{\rho_{i}}{L}u.
\]

The collision region, where the counterstreaming flows slow down as
they approach the mid-plane, is expected to undergo the following.
Initially, one of the rapidly growing electron-scale instabilities
sets in, depending on the velocity of the relative motion and thermal
velocities of the interpenetrating plasmas. The typical speed of the
relative motion is of the order of a several $10$km/s \citep{Sarkisov2005},
whereas the electron thermal velocity is considerably higher, up to
$10^{3}$km/s. The Buneman-type two-stream electron instability requires
$u>V_{Te},$and probably does not start. Even if it does, it will
quickly thermalize and mix up the electron populations before the
much heavier ions start responding to this instability. The modified
two-stream instability, in which electrons ExB drift in the wire direction,
has been considered \citep{MalkovSotnikov2018PhPl}. It should produce
a similar effect of electron preheating. After the saturation of these
fast instabilities, a slower instability of the counterstreaming ions
will develop at their (slower) time scale. A simple dispersion relation
for this instability, written down in the limit of monoenergetic counterstreaming
ion beams, has the following form

\begin{equation}
1+\chi_{e}-\frac{\omega_{pi}^{2}}{\left(\omega-ku\right)^{2}}-\frac{\omega_{pi}^{2}}{\left(\omega+ku\right)^{2}}=0\label{eq:DispEq1}
\end{equation}
The last two terms come from the counterstreaming ions, moving at
the bulk speeds, $\pm u$. The electron susceptibility, $\chi_{e},$
can be taken in an adiabatic approximation form

\begin{equation}
\chi_{e}=\frac{2}{k^{2}\lambda_{De}^{2}},\label{eq:AdElectr}
\end{equation}
which requires $\omega\ll k_{z}V_{Te}$. This requirement is consistent
with the electron preheating during the early stage of collision.
The electron contribution in eq.(\ref{eq:AdElectr}) to the dielectric
function corresponds to their density perturbations related to that
of the electrostatic potential as $n_{\text{e}}^{\prime}/n_{0}\approx e$$\phi^{\prime}/T_{\text{e}}$.
The coefficient '2' in the above expression stands for the two electron
populations that neutralize the respective ion beams of a number density
$n_{0}$. 

The role of the magnetic field is difficult to foresee, as it is of
opposite polarity in the inflowing coronas and may undergo a significant
turbulent reconnection/annihilation as the plasmas mix together. It
is possible that as the two plasmas expand adiabatically before they
collide, the regime of hot electrons, implied in eqs.(\ref{eq:DispEq1})
and (\ref{eq:AdElectr}), is established only after significant interpenetration
and heating of the coronas. During the initial collision phase, an
opposite regime of cold electrons, $k_{z}V_{Te}\ll\omega$, is more
likely to occur. Assuming that the unstable waves have a small wave
vector component along the magnetic field at this stage, $\chi_{e}$
takes a different form. Indeed, if the electrons are magnetized in
their motion across the local magnetic field, $\omega_{ci}\ll\omega\ll\omega_{ce}$,
instead of eq.(\ref{eq:AdElectr}) we can write
\begin{equation}
\chi_{e}=2\frac{\omega_{pe}^{2}}{\omega_{ce}^{2}}-2\frac{k_{z}^{2}}{k^{2}}\frac{\omega_{pe}^{2}}{\omega^{2}}\label{eq:MagnEl}
\end{equation}
The significance of this form of $\chi_{e}$ is a lower instability
threshold for certain wavenumbers. In this case, however, the instability
saturates more easily by electron heating. We will therefore override
the phase of initial electron heating and return to the adiabatic
regime, which we justify below. But, first, let us introduce the following
dimensionless parameters:

\[
\xi=\frac{k^{2}u^{2}}{\omega_{pi}^{2}},\,\,\,\,\,\,\,\,\,\,\mathcal{M}^{2}=\frac{u^{2}}{C_{s}^{2}}\equiv M\frac{u^{2}}{T_{e}}.
\]
Using this notation, eq.(\ref{eq:DispEq1}) rewrites

\[
1+\frac{2\mathcal{M}^{2}}{\xi}-\frac{1}{\left(\Omega-\sqrt{\xi}\right)^{2}}-\frac{1}{\left(\Omega+\sqrt{\xi}\right)^{2}}=0.
\]
One of its solutions describes an aperiodically growing mode, Fig.\ref{fig:Growth-rate-of}:

\begin{equation}
\frac{\omega^{2}}{\omega_{pi}^{2}}\equiv\Omega^{2}=-\frac{\xi}{\xi+2\mathcal{M}^{2}}\left(\sqrt{8\mathcal{M}^{2}+4\xi+1}-1-2\mathcal{M}^{2}-\xi\right)\label{eq:GrowthRate2stre}
\end{equation}
The aperiodic instability ($\omega^{2}<0$) develops when the term
in parentheses in eq.(\ref{eq:GrowthRate2stre}) is positive. The
condition for that can be written in the following short form:

\begin{equation}
\mathcal{M}^{2}+\frac{\xi}{2}<1\label{eq:InstConditMach}
\end{equation}
We observe that this condition can be met only in subsonic flows.
Such flows are expected in the plasma collision region, even if the
flows are initially highly supersonic, such as $u\gtrsim V_{Te}$.
In this case, as we discussed, the electrons should be heated via
Buneman instability. Likewise, the interpenetrating flows will also
be subsonic when standing termination shocks are formed on the two
sides of the collision zone. We should note here, though, that a situation
in which the flows are supersonic but not to the extent sufficient
for triggering the Buneman instability is also possible. In this case,
electrons can be preheated by other instabilities, such as the modified
two-stream instability, leading to the generation of lower-hybrid
waves. This case has already been considered in Ref.\citealp{MalkovSotnikov2018PhPl}.
We may conclude from the above considerations that at least one of
the discussed instabilities will result in electron preheating. We
will, therefore, continue our analysis of a slower, two-stream ion
instability under the assumption $\omega\ll k_{z}V_{Te}$.

The above ion-driven instability can be suppressed if ions in the
beam are also heated by generated waves or arrive in the collision
zone with a significant velocity dispersion, carried over from the
wire evaporation. We assume the finite ion thermal speed in each ion
beam, $V_{Ti}$, to examine this possibility. Eq.(\ref{eq:DispEq1})
generalizes to the following \citep{VVS_1961Usp} (see Appendix)

\begin{equation}
1+\frac{2}{k^{2}\lambda_{De}^{2}}-\frac{\omega_{pi}^{2}}{\left(\omega-ku\right)^{2}-k^{2}V_{Ti}^{2}}-\frac{\omega_{pi}^{2}}{\left(\omega+ku\right)^{2}-k^{2}V_{Ti}^{2}}=0\label{eq:DispEqWithPressure}
\end{equation}
Again, coefficient 2 in the electron contribution accounts for the
two electron components that neutralize the two counterstreaming ion
beams. These are represented by the last two terms on the l.h.s. of
the above equation. Using the dimensionless variables introduced earlier,
we rewrite this equation as follows

\begin{equation}
1+\frac{2}{k^{2}\lambda_{De}^{2}}-\frac{1}{\left(\Omega-\sqrt{\xi}\right)^{2}-k^{2}\lambda_{Di}^{2}}-\frac{1}{\left(\Omega+\sqrt{\xi}\right)^{2}-k^{2}\lambda_{Di}^{2}}=0\label{eq:DimLdispPressure}
\end{equation}
A compact expression for the solution can be obtained using the parameters
$\mathcal{M}$ and $\zeta\equiv2+\xi/\mathcal{M}^{2}$:

\begin{equation}
\Omega^{2}\equiv\frac{\omega^{2}}{\omega_{pi}^{2}}=\left(\zeta-2\right)\left[\mathcal{M}^{2}+\frac{1}{\zeta}+\frac{T_{i}}{T_{e}}-\frac{1}{\zeta}\sqrt{4\zeta\mathcal{M}^{2}\left(1+\frac{T_{i}}{T_{e}}\zeta\right)+1}\right]\label{eq:GroRa3D}
\end{equation}
For the onset of instability, the r.h.s of this expression should
be negative. This condition can be manipulated into the following
inequality

\[
0<\zeta\left(\mathcal{M}^{2}-\frac{T_{i}}{T_{e}}\right)<2,
\]
which, in turn, can be rewritten as the following constraint on the
Mach number of the counterstreaming flows:

\[
\frac{T_{i}}{T_{e}}<\mathcal{M}^{2}<\frac{T_{i}}{T_{e}}+\frac{1}{1+k^{2}C_{s}^{2}/2\omega_{pi}^{2}},
\]
or on their velocity:

\begin{equation}
\frac{T_{i}}{M}<u^{2}<\frac{T_{i}}{M}+\frac{C_{s}^{2}}{1+k^{2}C_{s}^{2}/2\omega_{pi}^{2}}.\label{eq:InstabWindow}
\end{equation}

\begin{wrapfigure}{o}{0.5\columnwidth}%
\includegraphics[clip,scale=0.47]{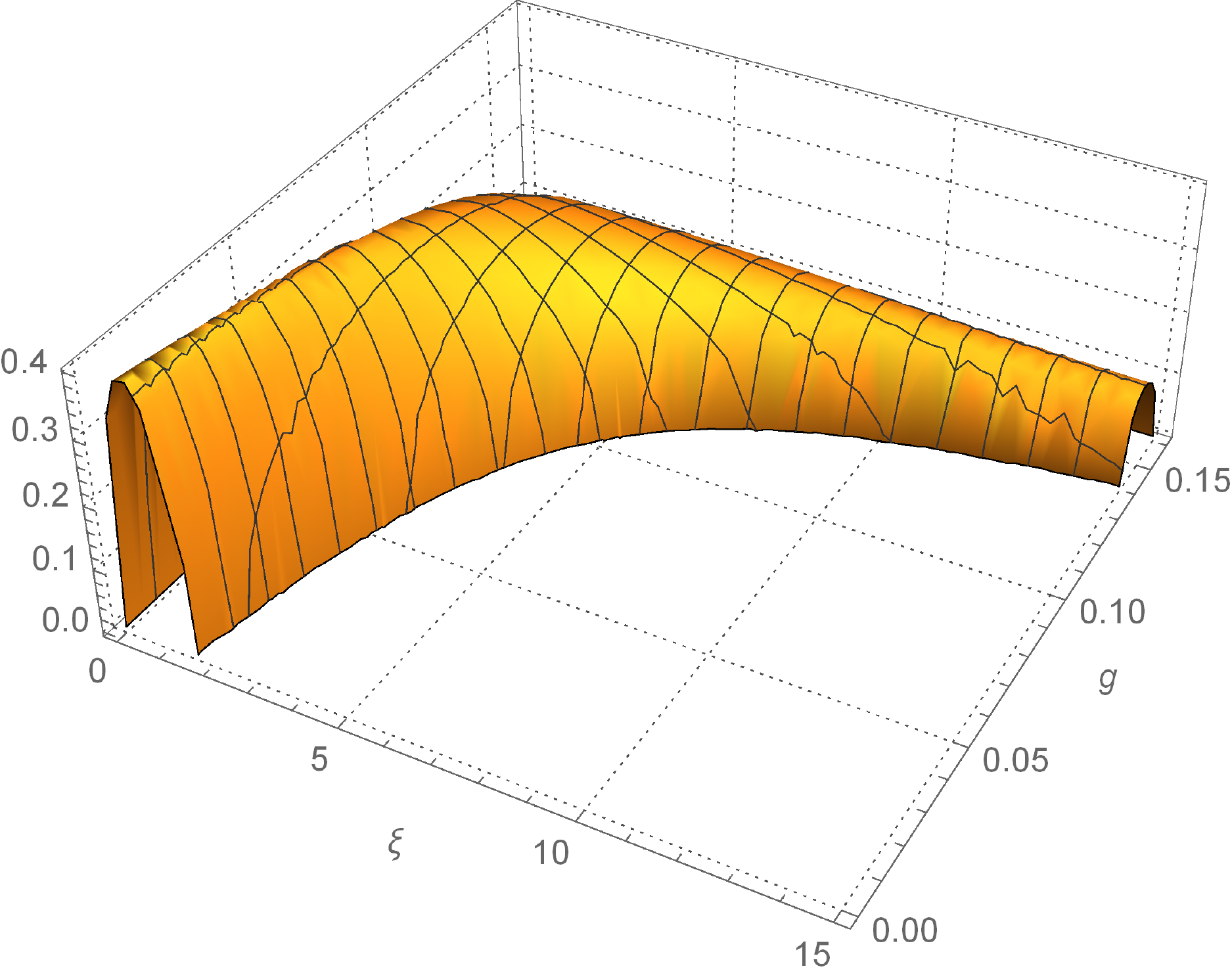}

\caption{Instability growth rate given by eq.(\ref{eq:GroRa3D}) for \emph{$\mathcal{M}=0.4$}
as a function of $\xi$ and $g\equiv T_{i}/T_{e}$.\label{fig:GrRate3D}}
\end{wrapfigure}

The growth rate given in eq.(\ref{fig:GrRate3D}) is shown in Fig.\ref{fig:GrRate3D}
as a function of $\xi$ and $T_{i}/T_{e}$ for $\mathcal{M}=0.4$.
The instability obviously, disfavors short waves, for which even insignificant
change of ion temperature or the flow velocity $u$, or both, may
produce a stabilizing effect. The electron heating, on the contrary,
widens the range of the flow velocities where the instability may
develop. Under these circumstances, it seems plausible to investigate
electron heating in more detail. Due to the low mass ratio, electrons
will be heated more efficiently than ions as the instability develops.
Electron heating broadens the velocity range of unstable modes and
impedes their stabilization, based on the above criterion.

\section{Electron Heating and Acceleration}

At least in a linear approximation, the instability considered above
excites electrostatic perturbations that are non-propagating in the
laboratory frame. Physically, the instability may be thought of as
intersecting backward-propagating ion-acoustic modes in each of the
two plasmas. They have phase velocities $\left|\omega_{\pm}/k\right|=u$,
where $\omega_{\pm}$ denote the wave frequencies in the frames of
the two counterstreaming plasmas. They resonate with each other in
the laboratory frame, in which they do not propagate. So, the mode
grows aperiodically. 

\subsection{Rapid Electron Preheating\label{subsec:Rapid-Electron-Preheating} }

The counterstreaming ion instability discussed in the previous section
has a growth rate $\gamma_{k}<\omega_{\text{pi}}$, so the electrons
respond adiabatically, as we assumed in deriving the dispersion relation,
that is $kV_{\text{Te}}\sim\omega_{\text{pi}}V_{\text{Te}}/u>\gamma_{k}$.
Representing then the fluctuating electric field as a superposition
of growing modes, $E=\sum_{k}E_{k}\left(t\right)\exp\left(i\boldsymbol{kr}\right)$,
we can write the following quasilinear equation for the heating process,
e.g., \citep{SagdGal69}

\begin{equation}
\frac{\partial f_{e}}{\partial t}=\frac{e^{2}}{m^{2}}\frac{\partial}{\partial\boldsymbol{v}}\sum_{k}\frac{\gamma_{k}\left|E_{k}\right|^{2}}{k^{2}v^{2}+\gamma_{k}^{2}}\frac{\partial f_{e}}{\partial\boldsymbol{v}}\label{eq:ElDiff1}
\end{equation}
This equation describes nonresonant interaction of electrons with
the unstable waves. Based on our assumption of electron adiabaticity,
$kV_{Te}>\gamma_{k}$, we can neglect the $\gamma_{k}^{2}$ in the
denominator of the above expression. Next, we introduce the electron
velocity diffusion coefficient and represent it in the form

\[
D_{e}=\frac{e^{2}}{2m^{2}}\frac{\partial}{\partial t}\sum_{k}\left|\phi_{k}\right|^{2}=\frac{e^{2}}{m^{2}}\sum_{k}\gamma_{k}\left|\phi_{k}\right|^{2},
\]
where we have denoted the spectral components of the wave electric
potential by $\left|\phi_{k}\right|=\left|E_{k}/k\right|$. Introducing
then a new variable

\[
\tau=\frac{e^{2}}{2m^{2}}\sum_{k}\left|\phi_{k}\left(t\right)\right|^{2},
\]
instead of $t$ and assuming first a one-dimensional spectrum of electrostatic
perturbations oriented along the flow, eq.(\ref{eq:ElDiff1}) can
be written in the following short form

\[
\frac{\partial f_{e}}{\partial\tau}=\frac{\partial}{\partial v}\frac{1}{v^{2}}\frac{\partial f_{e}}{\partial v}
\]
The last equation has a simple self-similar solution:
\[
f_{e}=\frac{C}{\tau^{1/4}}\exp\left(-\frac{v^{4}}{16\tau}\right)
\]
where C is a normalization constant. This solution describes reversible
electron heating that disappears together with the field oscillations.
The perceived electron ``temperature'', $T_{e}^{\text{*}}=2\sqrt{\tau}$,
in the above formula merely reflects the amplitude of coherent electron
oscillations in an ensemble of waves with random phases. 

In the case of three-dimensional isotropic fluctuations, the above
equation takes a slightly different form

\[
\frac{\partial f_{e}}{\partial\tau}=\frac{1}{v^{2}}\frac{\partial^{2}f_{e}}{\partial v^{2}}
\]
but its solution has a similar dependence on the particle velocity:

\[
f_{e}=\frac{C}{\tau^{3/4}}\exp\left(-\frac{v^{4}}{4\tau}\right),
\]
as well as the same physical meaning as in the one-dimensional case. 

While the adiabatic heating of electrons may impart significant amount
of kinetic energy, the energized electrons will cool down as soon
as the oscillations are damped or disappear otherwise, e.g., propagate
away from the excitation region. On the other hand, it is not difficult
to imagine a situation in which the entropy of electron gas increases,
thus rendering the heating process irreversible with no substantial
energy input. It might come about in two different ways or as a combination
thereof. In one way, the random but fixed phases of the waves begin
to evolve dynamically with a certain degree of stochasticity, which
should randomize the electron trajectories. In a second way, the electron
orbits are randomized independently of the oscillating field that
sets them in motion. Such a randomizing factor can be the scattering
of electrons on ions and neutrals in the shocked plasma layer. Since
we expect such scattering in the mixing layer, we include it in the
electron kinetics below, which will make the adiabatic heating irreversible.

\subsection{Stochastic Heating and Runaway Acceleration of Electrons in The Shocked
Plasma Layer \label{sec:Stochastic-Heating-and}}

Observations of enhanced emission near the flow stagnation region
in experiments, e.g., \citep{Sarkisov2005}, suggest that the shocked
plasma thermalizes most rapidly where the counterstreaming flows collide
head-on. Due to their rapid response to the instability, considered
earlier, the two counterstreaming electron distributions will couple
and form essentially one distribution with a zero average drift speed
across the layer. At the same time, ions will continue counterstreaming,
thus providing free energy for the instability considered in the preceding
section. The instability should, however, eventually saturate. The
most straightforward saturation mechanism is the convection of electrostatic
perturbations with plasma outflow from the instability region in the
layer direction. 

Meanwhile, the merger of electron distributions cannot be completed
everywhere in the mixing layer, especially at its edges. s. Therefore,
we first describe electrons that inflow into the mixing region from
its opposite sides as separate components and denote their distributions
as $f_{\pm}$, so that the full electron distribution is simply $f=f_{\text{+}}+f_{-}$.
The distributions $f_{\pm}$ enter the collision layer with drift
velocities $\pm u$, respectively. The equations for $f_{\pm}$ can
be written as follows 

\begin{equation}
\frac{\partial f_{\pm}}{\partial t}+\left(v\mu\pm u\right)\frac{\partial f_{\pm}}{\partial z}-\frac{e}{m}E\left(\mu\frac{\partial f_{\pm}}{\partial v}+\frac{1-\mu^{2}}{v}\frac{\partial f_{\pm}}{\partial\mu}\right)=\frac{1}{v^{2}}\frac{\partial}{\partial v}\left[v^{2}\nu_{\text{e}}\left(V_{\text{Te}}^{2}\frac{\partial f_{\pm}}{\partial v}+vf_{\pm}\right)\right]+\frac{\nu}{2}\frac{\partial}{\partial\mu}\left(1-\mu^{2}\right)\frac{\partial f_{\pm}}{\partial\mu}\label{eq:FPglobal}
\end{equation}
where we have denoted 

\begin{equation}
\nu_{\text{e,i}}=\frac{4\pi e^{4}N_{\text{e,i}}}{m^{2}v^{3}}\ln\left(\frac{mv^{2}\lambda_{\text{De}}}{e^{2}}\right)\label{eq:Nu-ei}
\end{equation}
and 

\begin{equation}
\nu=\nu_{\text{i}}+\nu_{\text{e}}\left(1-\frac{V_{\text{Te}}^{2}}{v^{2}}\right)\label{eq:NuiPlusNue}
\end{equation}

Below, explain the two equations for $f_{\pm}$ combined in eq.(\ref{eq:FPglobal})
for simplicity. Formally, they are written in separate reference frames
moving with velocities $\pm u$ in $z-$direction, so that the velocity
projection on $z$ direction is $v\mu\pm u$ in each of these equations.
The cosine of the particle pitch-angle to the shock normal ($z$-
direction), $\mu$, is introduced in a standard form $v_{z}=v\mu$,
where $v_{z}$ is the $z-$ component of electron velocity $\boldsymbol{v}$.
Apart from the terms with $\pm u$ on the lhs of eqs.(\ref{eq:FPglobal}),
the equations for $f_{\pm}$ are identical. Moreover, since $v$ is
at least $v\sim V_{\text{Te}}\gg V_{\text{Ti}}\sim u$, or even $v\gg V_{\text{Te}}$,
since we expect significant electron heating and acceleration by turbulent
electric field, $E$, we assume that $f_{\text{+}}\approx f_{-}$
inside the shocked layer. The role of the $\pm u$ term is then limited
to the edges of the layer where initially cold electrons enter the
layer from opposite sides. Likewise, the pitch-angle scattering operator
(the last term on the rhs) can be regarded as acting in the laboratory
rather than separately in each of the comoving reference frames, as
originally required in eq.(\ref{eq:FPglobal}). Except for $z-$dependence
(introduced straightforwardly), separation of the two groups of electrons
$f_{\pm}$, and an oscillatory rather than the constant electric field,
$E$, eq.(\ref{eq:FPglobal}) is similar to that used in studies of
electron runaway, e.g., \Citep{Gurevich61RunAway,KruskalBernstein64}.

\subsubsection{Equation for runaway electrons \label{subsec:Equation-for-runaway}}

Taking the above discussion into account, introducing pitch-angle
averaged quantities

\[
\left\langle \cdot\right\rangle =\frac{1}{2}\int_{-1}^{1}\left(\cdot\right)d\mu,
\]
and averaging eq.(\ref{eq:FPglobal}) over $\mu$, we arrive at the
following equation for $f=f_{\text{+}}+f_{-}$:

\begin{equation}
\frac{\partial\left\langle f\right\rangle }{\partial t}+\frac{1}{2}\left(\frac{eE}{mv^{2}}\frac{\partial}{\partial v}v^{2}-v\frac{\partial}{\partial z}\right)\left\langle \left(1-\mu^{2}\right)\frac{\partial f}{\partial\mu}\right\rangle =\frac{1}{v^{2}}\frac{\partial}{\partial v}\left[v^{2}\nu_{\text{e}}\left(V_{\text{Te}}^{2}\frac{\partial\left\langle f\right\rangle }{\partial v}+v\left\langle f\right\rangle \right)\right]+\left\langle S\right\rangle \label{eq:FPaver1}
\end{equation}
Here the term $\left\langle S\right\rangle $ arose from summing up
the fluxes $\pm uf_{\pm}$ through the boundaries of the layer. They
are not accounted for automatically when we add the equations for
$f_{\pm}$ together in eq.(\ref{eq:FPglobal}) and assume $f_{\text{+}}=f_{-}$.
The reason for this extra term is that $f_{\text{+}}$$\neq f_{-}$
near the boundaries, which we mentioned earlier. We may include in
the term $\left\langle S\right\rangle $ other transport channels,
such as particle convection along $x$ and y axes. To obtain an evolution
equation for $\left\langle f\right\rangle $, we need to express the
$\left\langle \left(1-\mu^{2}\right)\partial f/\partial\mu\right\rangle $
in the last equation through $\left\langle f\right\rangle $. This
can be done by applying Chapman-Enskog method to eq.(\ref{eq:FPglobal}).
This method will provide an expansion of equation for $\left\langle f\right\rangle $
in powers of $\overline{E^{2}}$- an average value of $E^{2}$ over
the distance and time that exceeds the mfp $\lambda=v/\nu$ and collision
time $1/\nu$, respectively. In particular, if the electric field
comes from the instability of counterstreaming ion beams considered
earlier, $E=\left(1/\sqrt{2}\right)\sum_{k}E_{k}\left(t\right)\exp\left(ikz\right)+c.c.$,
then $\overline{E^{2}}=\sum_{k}\left|E_{k}\right|^{2}$. 

The Chapman-Enskog method selects the zeroth order approximation that
makes the rhs (particle collision part) vanish. This is the assumption
for a system being close to thermal equilibrium. However, the collision
operator on the rhs of eq.(\ref{eq:FPglobal}) acts on the angular
part and the absolute value of particle velocity separately. Now we
consider the second term (angular part) on the rhs of eq.(\ref{eq:FPglobal})
to be dominant. Although it is formally of the same order as the first
term, it must vanish in the zeroth order by virtue of eq.(\ref{eq:FPaver1}),
as we can set $\left\langle S\right\rangle \approx0$ inside the shocked
plasma, as discussed earlier. Also note that for a Maxwellian distribution,
$f\propto\exp\left(-v^{2}/V_{\text{Te}}^{2}\right)$, the collision
term vanishes. As we shall see, for higher velocities the runaway
due the secular action of electric field dominates the thermal equilibtration
represented by the rhs in eq.(\ref{eq:FPaver1}). This dominance justifies
neglecting the first term on the rhs of eq.(\ref{eq:FPglobal}) for
the purpose of calculation of anisotropic correction to the equilibrium
state. 

Turning to the lhs of eq.(\ref{eq:FPglobal}) we note that the first
term ($\partial f/\partial t$) needs to be dropped to this order
of approximation as it is responsible for a short-time relaxation
of the anisotropic component, e.g. \citep{MS2015}. From eq.(\ref{eq:FPglobal}),
to the first order, we thus write

\[
\frac{\partial f}{\partial\mu}\approx\frac{1}{\nu}\left(\frac{eE}{m}\frac{\partial\left\langle f\right\rangle }{\partial v}-v\frac{\partial\left\langle f\right\rangle }{\partial z}\right)
\]
On substituting this expression into eq.(\ref{eq:FPaver1}), the latter
takes the following form 

\begin{equation}
\frac{\partial\left\langle f\right\rangle }{\partial t}=\frac{1}{v^{2}}\frac{\partial}{\partial v}v^{2}\left[\left(\nu_{\text{e}}V_{\text{Te}}^{2}+\frac{e^{2}\overline{E^{2}}}{3m^{2}\nu}\right)\frac{\partial\left\langle f\right\rangle }{\partial v}+\nu_{\text{e}}v\left\langle f\right\rangle \right]+\left\langle S\right\rangle +\frac{v^{2}}{3\nu}\frac{\partial^{2}\left\langle f\right\rangle }{\partial z^{2}}\label{eq:FPaver2}
\end{equation}
The last term on the rhs governs the diffusive losses from the shocked
layer. They might extend to the other coordinates, which can symbolically
be included into the $\left\langle S\right\rangle $-term, as we mentioned
earlier. The losses become important for large $v$, where they equate
the acceleration $\propto\overline{E^{2}}$. The acceleration occurs
in the plasma mixing layer that is bound by a pair of termination
shocks at $z=\pm l/2$. A plausible assumption about the distribution
of accelerated electrons in $z$- direction is that their density
vanishes at $z=\pm l/2$. The reason for this assumption is twofold.
First, the shocked layer has an elevated electrostatic potential that
slows down the inflowing ions in $z$ direction. Hence, this region
is a potential well for electrons, so they tend to have a higher concentration
toward the middle of this region. Second, the high-energy electrons,
for which the losses become significant, most likely never return
to the shocked layer if they cross one of its edges. Indeed, the flow
upstream of the termination shocks is relatively laminar. This means
that hot electrons accelerated in the mixing layer by the turbulent
electric field are not deflected strongly from their paths. Their
transport regime changes at the edge from diffusive to ballistic.
Depending on the electron energy and magnetic field strength, electrons
leave the layer by either following rectilinear trajectories or by
ExB drifting away from the flow stagnation region across the magnetic
field convected with the upstream flow. Therefore, we may use the
following expansion for $\left\langle f\right\rangle $ in $z$- direction 

\begin{equation}
\left\langle f\right\rangle =\sum_{n=1}^{\infty}f_{n}\left(t,v\right)\cos\left(\frac{\pi nz}{l}\right)\label{eq:FourExp-in-z}
\end{equation}
While this representation implies instantaneous losses of hot electrons
from the shocked layer as soon as they reach its boundaries at $z=\pm l/2$,
there is also an influx of cold electrons from the upstream. By keeping
only the leading term in the last Fourier expansion, eq.(\ref{eq:FPaver2})
can be rewritten as follows

\begin{equation}
\frac{\partial f_{1}}{\partial t}=\frac{1}{v^{2}}\frac{\partial}{\partial v}v^{2}\left[\left(\nu_{\text{e}}V_{\text{Te}}^{2}+\frac{e^{2}\overline{E^{2}}}{3m^{2}\nu}\right)\frac{\partial f_{1}}{\partial v}+\nu_{\text{e}}vf_{1}\right]+S_{1}-\frac{\pi^{2}v^{2}}{3l^{2}\nu}f_{1}\label{eq:FPaver3}
\end{equation}
Here, we have included the input of the $\left\langle S\left(z\right)\right\rangle $-term
in eq.(\ref{eq:FPaver2}) and denoted it by $S_{1}$. Note that both
$V_{\text{Te}}$ and $\overline{E^{2}}$ in the last equation, generally
speaking, depend on time. We start, however, from considering a steady
state solution of the above equation, assuming that these quantities
have already attained their time-asymptotic magnitudes.

\subsubsection{Steady State Fermi acceleration of Runaway Electrons\label{subsec:Steady-State-Solution}}

Since the source term $S_{1}$ in eq.(\ref{eq:FPaver3}) is associated
with the influx of relatively cold electrons into the shocked plasma
layer, it is plausible to place it at the origin in the velocity space.
Thus, we set $S_{1}=S\delta\left(v\right)/v^{2}$. Then, a steady
state solution for $v>0$ corresponds to a constant electron flux
$S$ between the source and the sink (the last term in eq.{[}\ref{eq:FPaver3}{]}
controls the particle losses) in velocity space:

\begin{equation}
S=-v^{2}\left[\left(\nu_{\text{e}}V_{\text{Te}}^{2}+\frac{e^{2}\overline{E^{2}}}{3m^{2}\nu}\right)\frac{\partial\left\langle f\right\rangle }{\partial v}+\nu_{\text{e}}v\left\langle f\right\rangle \right]=cosnt\label{eq:SfluxConst}
\end{equation}
\begin{figure}
\includegraphics[scale=0.48]{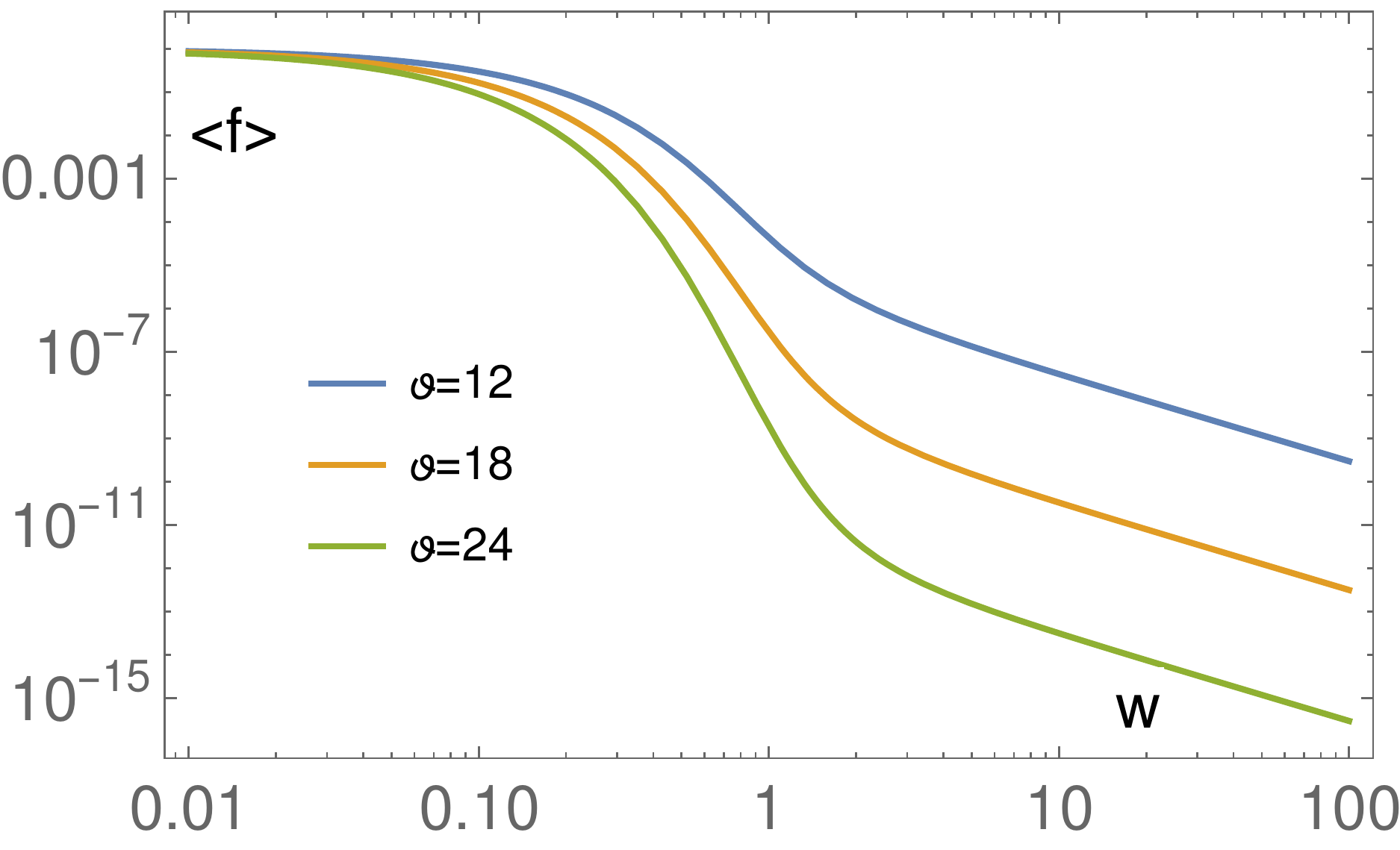}\includegraphics[scale=0.48]{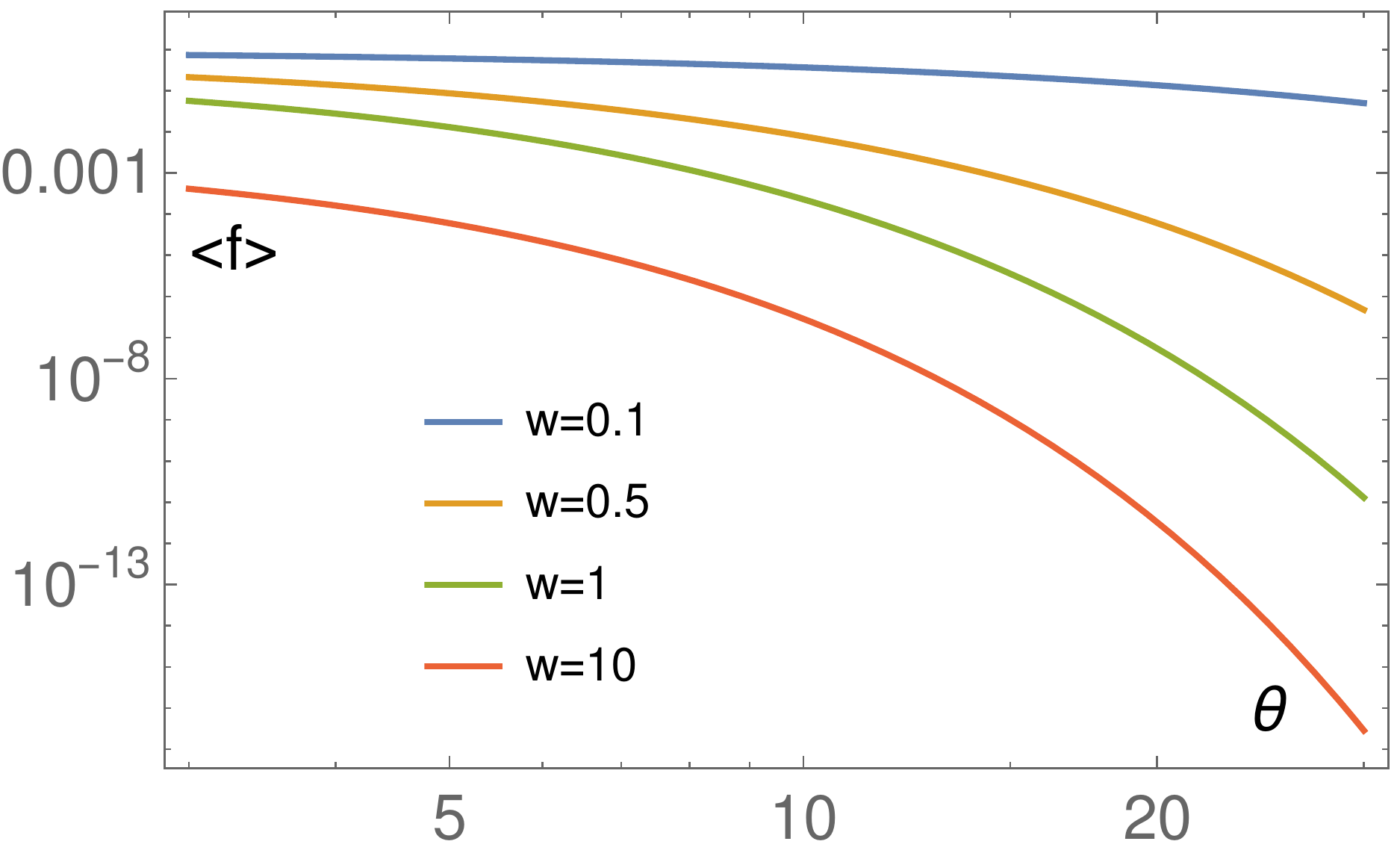}\caption{Left panel: steady state solutions of eq.(\ref{eq:SteadStateSp})
plotted for different $\vartheta$. To emphasize the development of
the runaway tail, the normalization constant $\sigma$ in each curve
is chosen to satisfy the condition $\left\langle f\left(0\right)\right\rangle =1$,
for all $\vartheta$ rather than fixing the total number of particles.
Right panel: the same solution shown as a function of $\vartheta$
for different particle energies, $w$. \label{fig:Steady-state-solutions}}
\end{figure}
This spectrum will cut off at sufficiently large $v$ due to losses,
which we have neglected for now because they grow very sharply at
high velocities (see eq.(\ref{eq:NuiPlusNue})) and are thus not important
in the velocity range considered. For relatively low values of $\overline{E^{2}}$
and not very large $v$ (for which $\nu$ and $\nu_{\text{e}}$ would
become small), the term with $E^{2}$ can be neglected and the solution
for $\left\langle f\right\rangle $ is close to a Maxwellian (because
$S=0$ as well for $v>0$). The electron runaway effect of the electric
field becomes important for $v\gg V_{\text{Te}}$, and we can substitute
$\nu\approx2\nu_{e}$ in eq.(\ref{eq:NuiPlusNue}). Furthermore, the
expression under the logarithm in eq.(\ref{eq:Nu-ei}) is typically
so large that the velocity variation under the logarithm can be neglected,
and $v$ can be set to a value $V_{\text{*}}$ at which the acceleration
and dynamical friction terms in eq.(\ref{eq:FPaver3}) equate, that
is 

\[
\frac{4\pi e^{4}N_{\text{e,i}}}{m^{2}v^{2}}\ln\left(\frac{mV_{\text{*}}^{2}\lambda_{\text{De}}}{e^{2}}\right)\sim\frac{e\sqrt{\overline{E^{2}}}}{m}
\]
By defining then the Dreicer's field, familiar from the theory of
runaway electrons \citep{Gurevich61RunAway,KruskalBernstein64} as
\[
E_{\text{D}}=\frac{4\pi e^{3}N_{\text{e,i}}}{T_{\text{e}}}\ln\left(\frac{mV_{\text{*}}^{2}\lambda_{\text{De}}}{e^{2}}\right),
\]
the critical velocity $V_{\text{*}}$ above which the runaway process
dominates dynamical friction can be expressed through $E_{\text{D}}$
as follows: 

\begin{equation}
V_{\text{*}}=V_{\text{Te}}\frac{E_{\text{D}}^{1/2}}{\left(\overline{E^{2}}\right)^{1/4}}\label{eq:Vcrit}
\end{equation}
It is now convenient to use dimensionless electron energy $w=v^{2}/6^{1/3}V_{\text{Te}}^{2/3}V_{\text{*}}^{4/3}$
and field strength parameter 
\begin{equation}
\vartheta=6^{1/3}V_{\text{*}}^{4/3}/2V_{\text{Te}}^{4/3}.\label{eq:thetaDef}
\end{equation}
Using these quantities, eq.(\ref{eq:SfluxConst}) can be rewritten
as follows

\begin{equation}
\left(1+w^{3}\right)\frac{\partial\left\langle f\right\rangle }{\partial w}+\vartheta\left\langle f\right\rangle =-S\vartheta\frac{m^{2}}{eT_{\text{e}}E_{\text{D}}}\equiv-\sigma\label{eq:SteadStateSp}
\end{equation}
This equation can be solved in quadratures:

\begin{equation}
\left\langle f\right\rangle =\sigma\exp\left[-\frac{\vartheta}{\sqrt{3}}\tan^{-1}\left(\frac{2w-1}{\sqrt{3}}\right)\right]\frac{\left(1-w+w^{2}\right)^{\vartheta/6}}{\left(1+w\right)^{\vartheta/3}}\int_{w}^{\infty}\frac{\left(1+w\right)^{\vartheta/3-1}dw}{\left(1-w+w^{2}\right)^{\vartheta/6+1}}\exp\left[\frac{\vartheta}{\sqrt{3}}\tan^{-1}\left(\frac{2w-1}{\sqrt{3}}\right)\right]\label{eq:SteadyStateSolInt}
\end{equation}
Fig.\ref{fig:Steady-state-solutions} (left panel) shows the above
solution for different values of field-strength parameter $\vartheta$.
It is seen that for strong electric fields, $\vartheta\sim1$, the
power-law tail, $\left\langle f\right\rangle \propto w^{-2}\propto v^{-4}$,
grows immediately from the thermal velocity. For weak fields ($\vartheta\gg1)$
the Maxwellian core is distinct from the power-law tail of runaway
electrons. On the right panel of Fig.\ref{fig:Steady-state-solutions},
the electron distribution function $\left\langle f\right\rangle $
is shown as a function of $\vartheta$ for several values of the fixed
electron energies. 

The asymptotic power-law energy spectrum $\left\langle f\right\rangle \propto w^{-2}$
may be interpreted as a classical Fermi acceleration at work. Fermi
\citep{Fermi49} showed that under competing particle acceleration-loss
events, a power-law energy spectrum, $w^{-q}$, is naturally established.
Its index $q$, which is in our case, $q=2$, is determined by the
ratio of particle confinement time, $\tau_{\text{loss}}$, and its
acceleration time, $\tau_{\text{acc}}$. Fermi originally devised
this mechanism for galactic cosmic rays, in which case $\tau_{\text{loss}}$
is an average time spent by a CR particle in the galaxy before it
escapes into the intergalactic space. The acceleration time, in turn,
is simply the energy e-folding time, $dE/dt=E/\tau_{\text{acc}}$,
where $E$ the particle energy. According to Fermi, the power-law
index, $q$, in the particle distribution, $f\propto E^{-q}$, can
be calculated as follows:

\[
q=1+\frac{\tau_{\text{acc}}}{\tau_{\text{loss}}}
\]
The Fermi acceleration has been recognized later as a general mechanism
pertinent to many other systems where a nonthermal spectrum emerges
from the balance between the particle energy gain and its loss or
particle escape. In our case, accelerated particles loose energy as
they scatter on slowly moving particles (thermal electrons, ions,
and neutrals). Between each scattering they gain energy from the electric
field. So, the energy gain and loss occur during the same time interval,
$\tau_{\text{loss}}=\tau_{\text{acc}}$. The result $q=2$ immediately
follows from the Fermi's celebrated formula. 

The asymptotic $q=2$ acceleration regime is pertinent to high-energy
particles, for which the contribution of thermal core into the flux
$S=const$ in eq.(\ref{eq:SfluxConst}) can be neglected (all terms
other than the acceleration term, $\propto\overline{E^{2}}$). The
acceleration term, however, includes both the actual energy gain between
two consecutive collisions and the particle scattering that results
from these collisions. This conclusion can be inferred from the derivation
of eq.(\ref{eq:FPaver3}) in Sec.\ref{subsec:Equation-for-runaway}.
The full steady state solution in eq.(\ref{eq:SteadyStateSolInt})
does include the thermal core part of the spectrum and it has a somewhat
complex form. For the high-energy particles, the asymptotic spectrum
can be most easily obtained from the relation $w^{3}\partial\left\langle f\right\rangle /\partial w=-\sigma=const$,
according to the insignificant role of the thermal core for particles
with $w\gg1$.

\subsubsection{Time-dependent runaway\label{subsec:Time-dependent-runaway}}

It is convenient to study time-dependent solutions of eq.(\ref{eq:FPaver3})
using a new variable $U$ instead of the particle energy $w$ or velocity
$v$. These variables are related in the following way: 

\[
U=w^{3/2}=\frac{v^{3}}{\sqrt{6}V_{\text{Te}}V_{\text{*}}^{2}}.
\]
By measuring time $t$ in the units of electron acceleration time,
$\tau_{\text{a}}$ ($t/\tau_{\text{a}}\to t$), and introducing the
particle loss parameter, $\lambda$, owing to their diffusion in $z-$
direction, 

\[
\tau_{\text{a}}^{-1}=6^{1/6}\frac{e}{m}E_{D}\frac{V_{\text{Te}}^{7/3}}{V_{\text{*}}^{10/3}},\;\;\;\;\;\lambda=\frac{3^{4/3}m^{2}}{4E_{D}^{2}e^{2}}\frac{V_{\text{*}}^{20/3}}{l^{2}V_{Te}^{8/3}}
\]
eq.(\ref{eq:FPaver3}) rewrites as follows

\begin{equation}
\frac{\partial f_{1}}{\partial t}=\frac{\partial}{\partial U}\left[\frac{3}{2}U^{1/3}\left(1+U^{2}\right)\frac{\partial f_{1}}{\partial U}+\vartheta f_{1}\right]+\tau_{a}S_{1}-\lambda U^{5/3}f_{1}\label{eq:FinalFor-f1}
\end{equation}
This equation contains two parameters, $\vartheta$ and $\lambda$;
these are the rates of collisional energy losses of runaway electrons
and their escape through the boundaries of the mixing layer, respectively.
Besides, the parameter $\vartheta$ determines the thermal core temperature
that establishes through a balance between dynamical friction (term
proportional to $\vartheta$) and velocity diffusion (term proportional
to $1+U^{2}$, where the first term in the sum corresponds to the
thermal velocity diffusion due to Column collisions, while $U^{2}$
stands for the anomalous velocity diffusion). The part of the velocity
diffusion $\propto U^{2}$ stems from the electric field generated
by the ion beam instability discussed in Sec.\ref{sec:Two-Stream-Instability-in}. 

According to our discussion in the preceding section, the source supplies
particles at lower energies, that is at $U\ll1$. To keep the treatment
simple, we therefore draw equivalence between the source term and
the boundary condition at the lower end of the integration domain.
In other words, instead of specifying the source profile $S_{1}\left(U\right),$
we keep the value $f\left(U_{\text{min}}\right)$ fixed as a boundary
condition. It is clear that acceleration is negligible near this end
as long as $U_{\text{min}}\ll1$. Also, to circumvent an insignificant
problem with the algebraic singularity, $U^{1/3}$, at $U=0$ in the
above equation, we set $U_{\text{min }}=0.1$ instead of $U_{\text{min }}=0$.
This simplification is justified by the upper end of the integration
interval being at $U_{\text{max}}\gg1$. We then impose $f_{1}\left(U_{\text{min}}\right)=1$
and $f_{1}\left(U_{\text{max}}\right)=0$ as a complete set of boundary
conditions for eq.(\ref{eq:FinalFor-f1}). The latter condition implies
that particles accelerated to $U_{\text{max}}$ escape the mixing
layer regardless of the value of $\lambda$, but the effect of finite
$\lambda$ is still present in the solution. 

\begin{figure}
\includegraphics[scale=0.55]{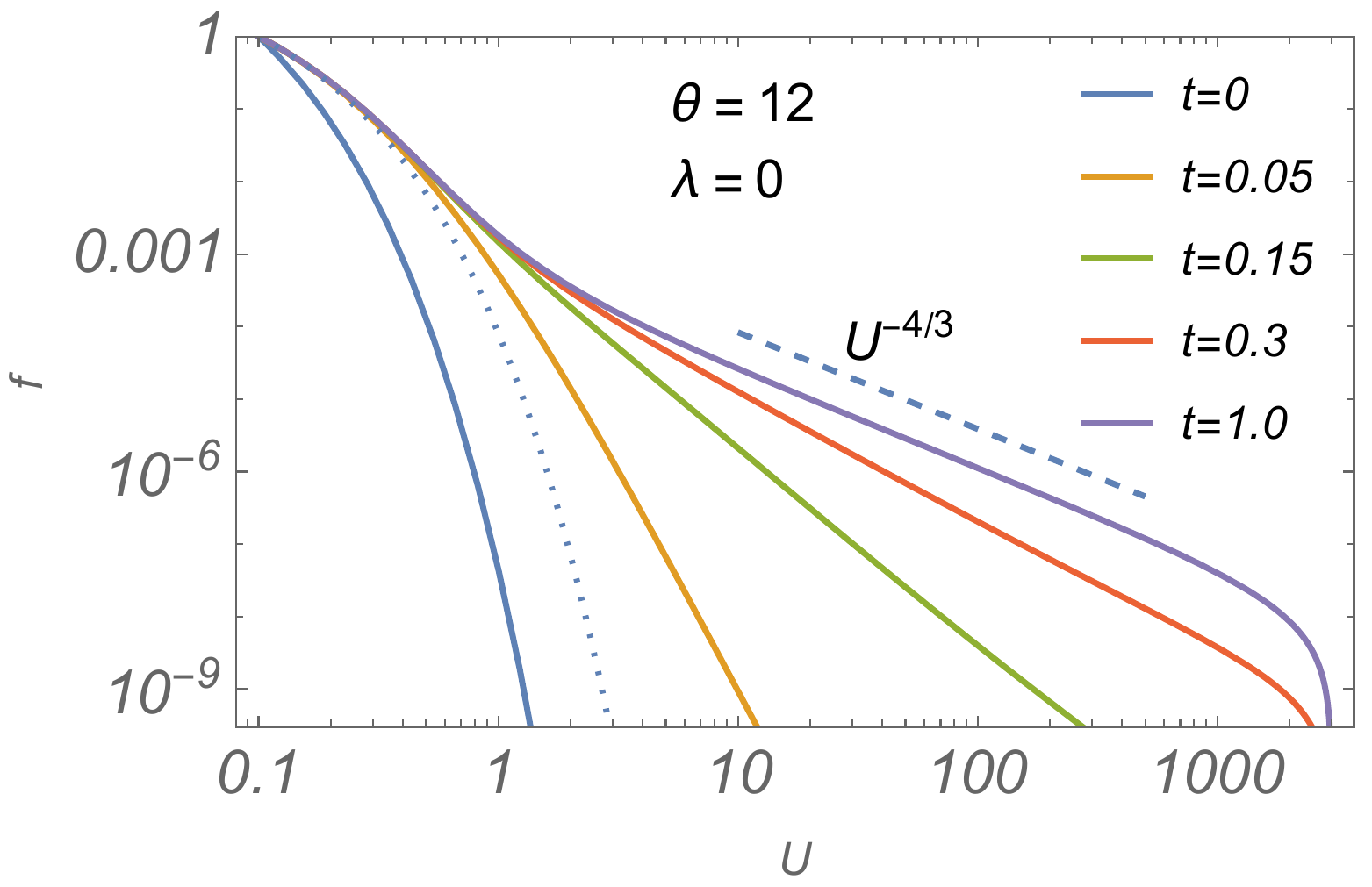}\includegraphics[scale=0.55]{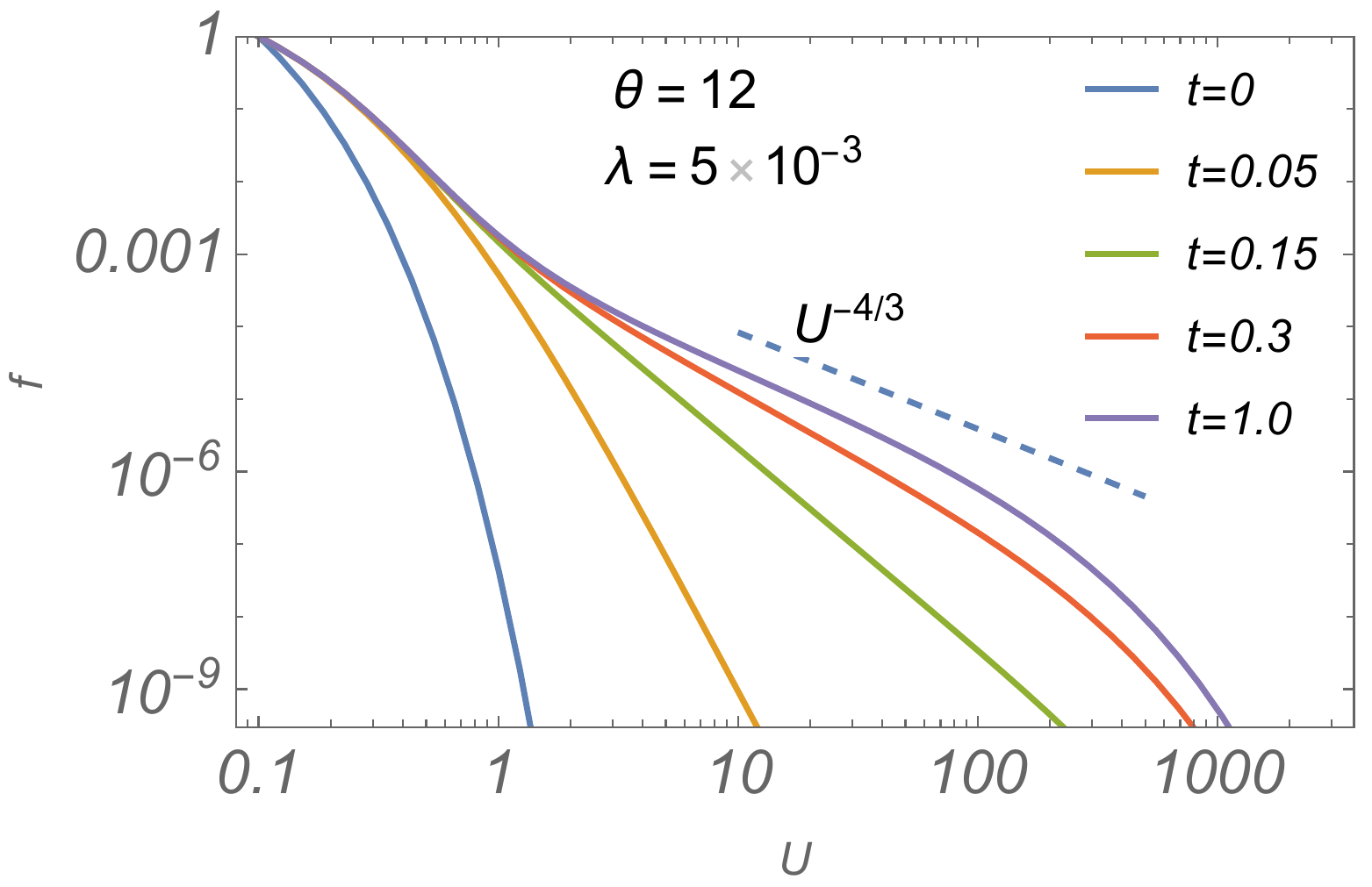}

\includegraphics[scale=0.55]{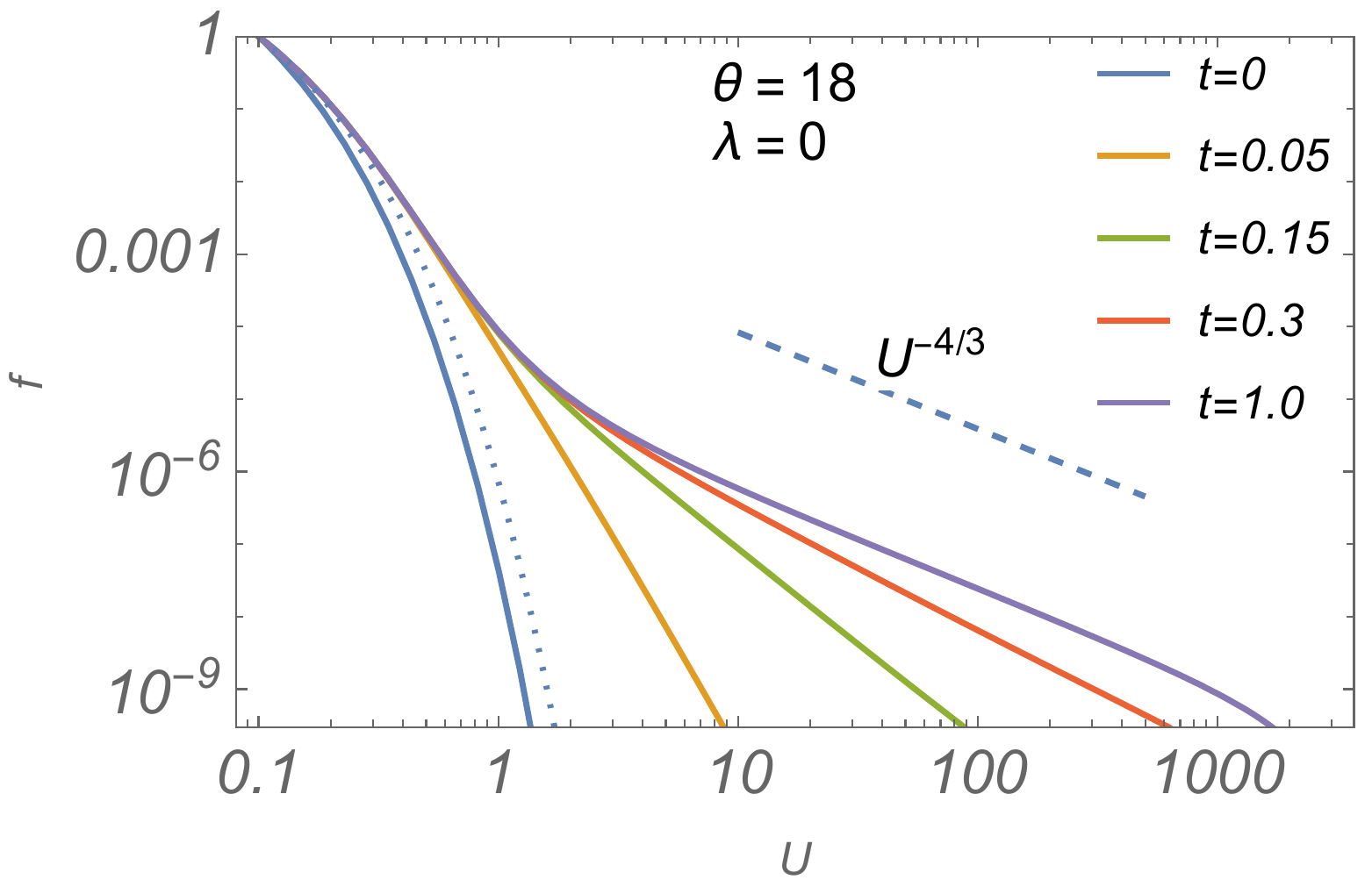}\includegraphics[scale=0.55]{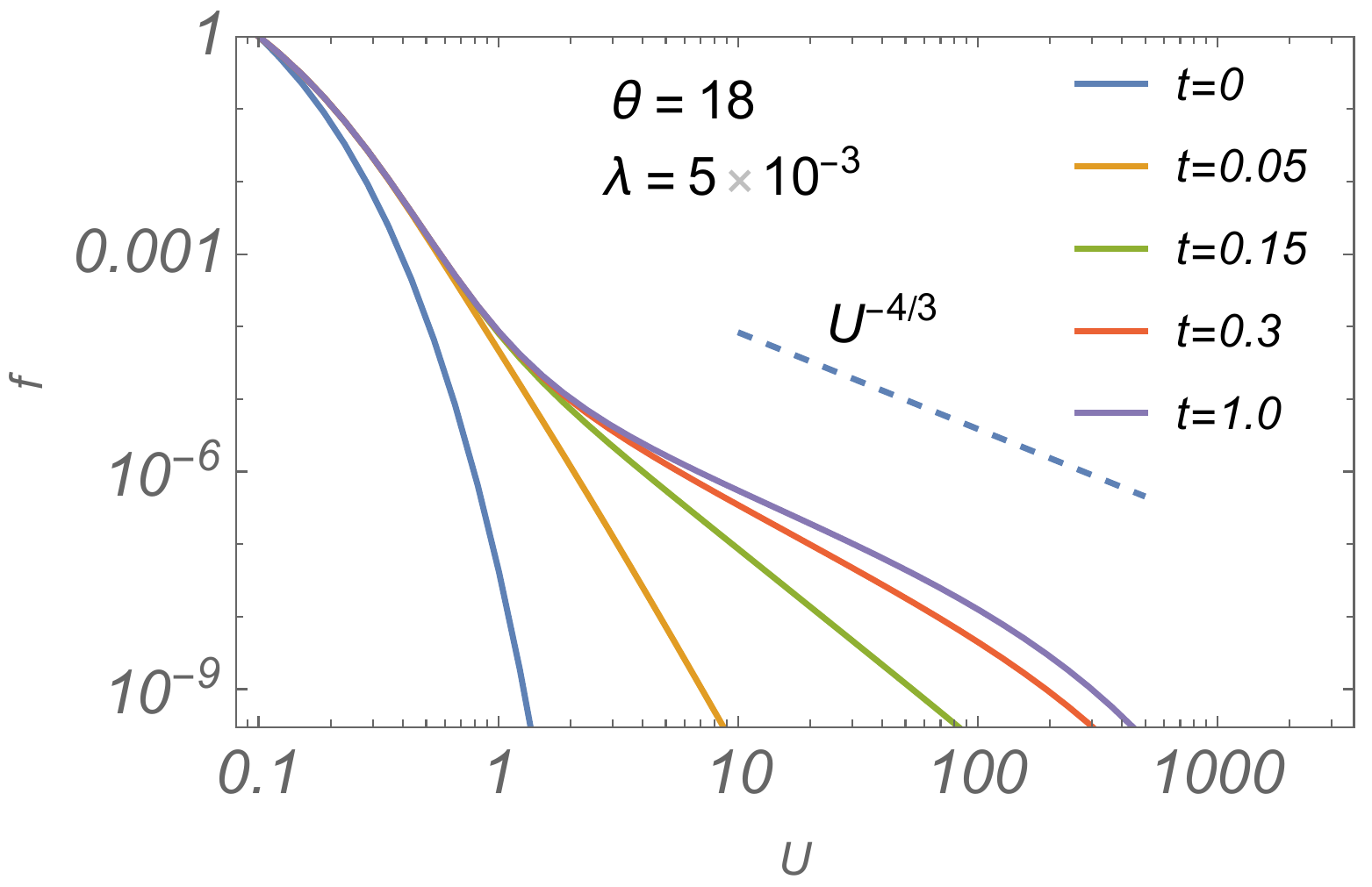}

\caption{Time evolution of the spectrum of runaway electrons, as described
by eq.(\ref{eq:FinalFor-f1}), starting from a relatively cold initial
distribution with $U_{\text{th}}=0.01$. Four combinations of equation
parameters $\vartheta$ and $\lambda$ are shown in each plot. The
left two panels show the loss-free solutions, $\lambda=0$, except
for a particle instantaneous escape upon reaching the maximum energy
at $U_{\text{max}}.$ \label{fig:Time-evolution-of}}
\end{figure}

Fig.\ref{fig:Time-evolution-of} illustrates the electron runaway
acceleration, as it evolves in time, shown for four combinations of
parameters $\vartheta$ and $\lambda$. All four runs start from a
relatively cold Maxwellian (if expressed in terms of particle velocity
rather than $U$):

\begin{equation}
f_{1}(0,U)=\exp\left[-\frac{U^{2/3}-U_{\text{min}}^{2/3}}{U_{\text{th}}^{2/3}}\right]\label{eq:ICthermal}
\end{equation}
where $U_{\text{th }}$ characterizes the initial electron temperature.
Although the electron adiabatic preheating, as described in Sec.\ref{subsec:Rapid-Electron-Preheating},
should result in a steeper decay of electron distribution, we assume
the above Maxwellian form. The reason for that choice is twofold.
First, the initial distribution is too narrow to significantly affect
its evolution beyond a short initial phase. Second, the collisional
relaxation of any initial distribution localized at $U\ll1$ dominates
the evolution of $f_{1}\left(t,U\right)$ at early times. It quickly
relaxes to $f_{1}\propto\exp\left(-\vartheta U^{2/3}\right)$, as
can be seen from eq.(\ref{eq:FinalFor-f1}). This initial dynamics
is seen in the two left panels of Fig.\ref{fig:Time-evolution-of}.
Indeed, for the chosen two values of dynamical friction parameter,
$\vartheta=12$ and $\vartheta=18$, the electron distribution broadens
to the respective Maxwellian, $\exp\left(-\vartheta U^{2/3}\right)$
at $U\lesssim1$ very quickly and proceeds then to higher energies.
At $U\gtrsim1$, the particle distribution develops a suprathermal
tail that also very quickly ``forgets'' the initial velocity distribution. 

While the initial distribution of electrons is not significant for
the development of the suprathermal tail, the tail amplitude is very
sensitive to the dynamical friction parameter, $\vartheta$ (see also
Fig.\ref{fig:Steady-state-solutions}). As seen from Fig.\ref{fig:Time-evolution-of},
for a 50\% increase in $\vartheta$ from the top panel to the bottom
ones, the tail height drops by about two orders of magnitude. This
is similar to the conventional runaway process (in a constant electric
field) as the suprathermal tail is detached from the thermal Maxwellian
at a critical velocity, $V_{\text{*}}$, which in our case is defined
in eq.(\ref{eq:Vcrit}) and is related to the dynamical friction,
$\vartheta$. Because of the exponential decay of Maxwellian distribution
at $v\sim V_{\text{*}}$, small variations of $V_{\text{*}}$ result
in large changes in the tail height, connected to the Maxwellian at
$v\approx V_{\text{*}}$. 

\section{Conclusions and Discussion\label{sec:Conclusions-and-Discussion}}

In this paper, we have studied microscopic processes occurring in
a mixing layer between two colliding plasma clouds. Our major findings
are as follows:
\begin{enumerate}
\item Initial relative velocity of interpenetrating plasmas needs to be
in the range between the ion thermal velocity, $V_{\text{Ti}}=\sqrt{T_{\text{i}}/M}$,
and ion-sound velocity, $C_{\text{s}}=\sqrt{T_{\text{e}}/M}$, for
triggering a strong anomalous coupling of the plasma streams that
is supported by the two-stream instability. 
\item Subsequent nonresonant (reversible) heating of electrons then broadens
the range of unstable wavenumbers.
\item Inclusion of electron-ion and electron-neutral collisions makes the
electron energization irreversible.
\item In addition to the heating of an electron core distribution, a suprathermal
power-law tail is produced in wave-particle interactions. It decays
with energy as $w^{-2}$, thus showing a notable analogy with the
stochastic Fermi acceleration.
\item The amplitude of the power-law electron tail strongly depends on the
ratio of the turbulent electric field to the electron collision frequency.
\end{enumerate}
While the instability condition in (1) may appear to be restrictive,
the following arguments can be advanced in its favor. Since each of
the colliding plasmas undergoes a transition from supersonic to subsonic
speed in the collision region, the instability condition is likely
to be met at least in a limited part of the collision region. The
subsequent electron heating will then broaden the unstable range of
relative velocity. 

Under favorable instability conditions, specified above and in more
detail in Sec.\ref{sec:Two-Stream-Instability-in}, a significant
fraction of kinetic energy of counterstreaming corona ions is likely
to be converted in ion-acoustic waves. When also the electron-ion
and electron-neutral collision frequencies are limited by the condition
$\vartheta=6^{1/3}V_{\text{*}}^{4/3}/2V_{\text{Te}}^{4/3}\lesssim10$
(here $V_{\text{*}}$, eq.{[}\ref{eq:Vcrit}{]}, is a critical electron
velocity beyond which they enter a runaway regime), the wave energy
is efficiently transferred to thermal and nonthermal electrons. The
anomalous (wave-driven) electron heating and acceleration will likely
to have a defining impact on the next phase of plasma mixing. 

The next phase starts when heavier wire-core materials, which follow
the light corona plasmas, begin to interpenetrate. As the materials
are largely in a neutral or weakly ionized state at the moment of
touching each other, the presence of hot and suprathermal coronal
electrons in the mixing layer will affect the interpenetration process
by primarily ionizing the neutrals and producing line emission. Although
this phase of the plasma collision was out of the scope of this paper,
focused on the coronal phase, it is worthwhile to briefly discuss
possible scenarios that may be derived from the presence of hot and
suprathermal electrons. 

As we mentioned, energetic electrons will ionize counterstreaming
wire-core atoms, thus partially converting them into counterstreaming
ion beams. The plasma stability analysis then becomes similar to that
of the coronal collision studied earlier. However, several differences
are evident. First, the beams will likely become multi-charged as
the ionization potential of wire material (e.g., aluminum, tungsten)
grows significantly with the charge number. Second, as the energetic
electron population left behind after the coronal phase of plasma
collision has a much lower concentration than the wire-core atoms,
the ionization is likely to occur in an avalanche regime \citep{GurevichUFN2001}.
The neutral gas electric breakdown supported by the oscillatory electric
field generated by the two-stream instability of freshly ionized beams
may then naturally explain the protracted line emission observed in
experiments with double-wire electric explosion \citep{Sarkisov2005}.
Note, that this phenomenon is not observed in a single-wire experimental
setup. Another important aspect of these experiments is that the voltage
collapse in the vacuum chamber occurs when the light emission starts.
Therefore, the electric discharge is not driven by the external electric
field, which also speaks to an internal mechanism of the observed
glow in the mixing layer. 

Strong radiation from the mixing layer suggests that its bounding
shocks may be radiative. The removal of the shocked plasma internal
energy by line emission softens the equation of state thus driving
the adiabatic index to $\gamma\approx1$, which makes the shocks more
compressive. The plasma layer between them becomes thinner. Our hydrodynamic
simulation \citep{Sotnikov2020} of colliding concentric wire outflow
have shown that the low value of $\gamma$$\approx1$ is required
to keep the mixing layer thin enough and not to expand under the internal
pressure of shocked plasma. The layer remains way inside the gap between
the original wire positions, as observed in experiments \citep{Sarkisov2005}.

\begin{acknowledgments}
This work was supported by the Air Force Office of Scientific Research under 
the grant LRIR No. 19RYCOR062. Public release approval record: AFRL-2021-4552
\end{acknowledgments}

\appendix

\section{Dispersion Equation for Interpenetrating Plasma Flows}

Thermal corrections to the plasma dispersion function are often given
in an additive form. It is more restrictive to the magnitude of these
corrections than those we used in eq.(\ref{eq:DispEqWithPressure}).
As a reference, we provide below its simple derivation, following
the work \citep{VVS_1961Usp}. The equations of motion for each
of the ion beams moving with the bulk velocities $\pm u,$ with the
respective perturbation, $u_{\pm}^{\prime},$ is

\begin{equation}
\frac{\partial u^{\prime}}{\partial t}\pm u\frac{\partial u^{\prime}}{\partial x}=-\frac{1}{Mn_{i}}\frac{\partial p^{\prime}}{\partial x}-\frac{e}{M}\frac{\partial\phi}{\partial x}\label{eq:ApEqOfMotion}
\end{equation}
To lighten the notation, we have omitted the subscripts $\pm$ in
$u^{\prime}$, $p^{\prime}$, $n^{\prime}$ for either beam, as it
should not cause any confusion. The respective equations for $u_{\pm}^{\prime}$,
etc., are marked by the sign in front of the beam velocity, $\pm u$.
In the above equations, we have denoted the pressure perturbation
as $p^{\prime}$, while $n_{i}$ below denotes the unperturbed ion
density of each beam and $\phi$ is the perturbation of electrostatic
potential. The continuity equation for the density perturbation, $n_{i}^{\prime}$

\begin{equation}
\frac{\partial n_{i}^{\prime}}{\partial t}\pm u\frac{\partial n_{i}^{\prime}}{\partial x}+n_{i}\frac{\partial u^{\prime}}{\partial x}=0\label{eq:ApContEq}
\end{equation}
while the Poisson equation

\begin{equation}
\frac{\partial^{2}\phi}{\partial x^{2}}=4\pi e\sum_{\pm}\left(n_{e}^{\prime}-n_{i}^{\prime}\right)\label{eq:ApPoisEq}
\end{equation}
where $n_{e}^{\prime}$ denotes the total electron density perturbation,
while the sum is taken over the two ion beams and the respective electron
contributions that are initially moving at the same speed as the ions.
We relate the ion pressure perturbation with that of their density,
assuming a constant temperature, $T_{i},$ $p^{\prime}=T_{i}n_{i}^{\prime},$while
a Boltzmann distribution for each electron component simply yields
$n_{e}^{\prime}/n_{i}=e\phi/T_{e}.$ Assuming that the perturbed quantities
depend on $x$ and $t$ as $\exp\left(ikx-i\omega t\right)$, expressing
$n_{i}^{\prime}$ through $\phi$ using eqs.(\ref{eq:ApEqOfMotion}-\ref{eq:ApContEq}),
and substituting in eq.(\ref{eq:ApPoisEq}), we arrive at eq.(\ref{eq:DispEqWithPressure})

\bibliographystyle{prtec}
\bibliography{}

\end{document}